\documentclass[a4paper]{spie}  %>>> use for US letter paper
\usepackage[]{graphicx}
\usepackage{hyperref}
\usepackage{aas_macros}
\usepackage{amsmath}
\usepackage{amsfonts}

\title{Approximate nonnegative matrix factorization algorithm for the analysis of angular differential imaging data} 

\author[a,c]{Arcidiacono C.} 
\author[b]{Simoncini V.}

\affil[a]{INAF -- Osservatorio Astrofisico e scienza dello Spazio di Bologna, Via P. Gobetti 93/3, 40129 Bologna, Italy}
\affil[b]{Universit\`a degli Studi di Bologna, Dipartimento di Matematica, Piazza di Porta S. Donato 5, 40126 Bologna, Italy}
\affil[c]{ADONI -- Laboratorio Nazionale di Ottica Adattiva, Italy}
\authorinfo{Further author information: (Send correspondence to Carmelo Arcidiacono)\\C.A.: E-mail: carmelo.arcidiacono@inaf.it, Telephone:  +39 051 6357 316}
\begin{document} 
  \maketitle 
	
%%%%%%%%%%%%%%%%%%%%%%%%%%%%%%%%%%%%%%%%%%%%%%%%%%%%%%%%%%%%% 
\begin{abstract}
The angular differential imaging (ADI) is used to improve contrast in high resolution astronomical imaging. An example is the direct imaging of exoplanet in camera fed by Extreme Adaptive Optics. The subtraction of the main dazzling object to observe the faint companion was improved using Principal Component Analysis (PCA). It factorizes the positive astronomical frames into positive and negative components. On the contrary, the Nonnegative Matrix Factorization (NMF) uses only positive components, mimicking the actual composition of the long exposure images. 
\end{abstract}

%>>>> Include a list of keywords after the abstract 


%%%%%%%%%%%%%%%%%%%%%%%%%%%%%%%%%%%%%%%%%%%%%%%%%%%%%%%%%%%%%
\section{INTRODUCTION}
\label{sec:intro}  % \label{} allows reference to this section
The front-line of the technological advancement for astronomical imaging can be reduced to the achievement of a better resolution and a better contrast.
Larger and larger telescopes are being deployed in space or built on the ground.
In both cases the major aim is to reduce the fundamental limit given by the diffraction, enhancing both the theoretical resolution and contrast.
High contrast imaging in astronomy is a key tool for most of the science cases of future Extremely Large Telescopes\cite{2013JApA...34...81S,2007Msngr.127...11G,GMT}. In the following, we study the case of high contrast technique such the Angular Differential Imaging (see section \ref{sec:adi}) used to exploit the power of extreme adaptive optics system. In this framework we propose the use of Nonnegative Matrix Factorization algorithms\cite{Paatero:Environmetrics94,lee99} to improve the data reduction, see section \ref{sec:nmf}.
\section{Angular Differential Imaging}
\label{sec:adi}
\begin{figure}
   \begin{center}
   \begin{tabular}{c}
   \includegraphics[width=10cm]{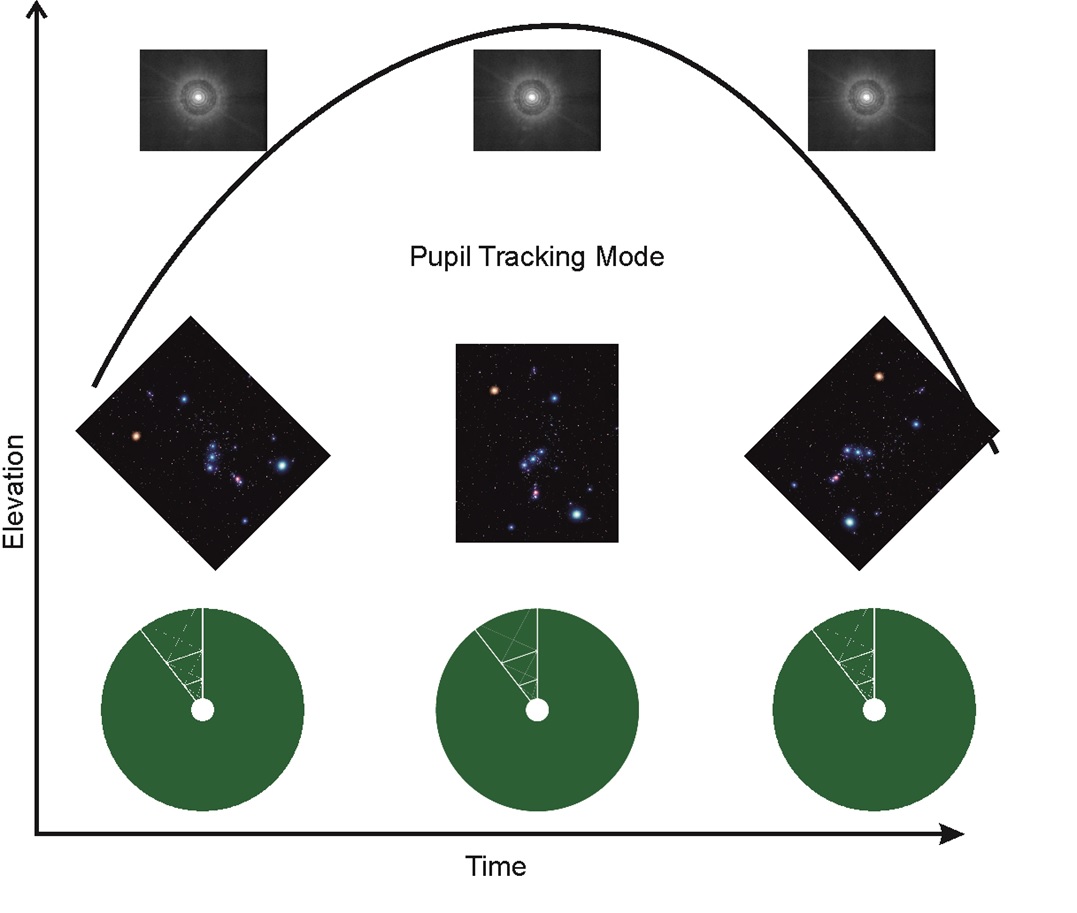}\\
   \end{tabular}
  \end{center}
   \caption[example] 
%>>>> use \label inside caption to get Fig. number with \ref{}
   { \label{fig:puptrack} 
On Alt-Az telescope field rotates as the object crosses the sky. Typically cameras are counter-rotated, in the ADI are not and the technique takes advantage of the rotation of the object with respect to the PSF fixed to the telescope.}
   \end{figure}

The angular differential imaging\cite{2006ApJ...641..556M} (ADI) is a data reduction and data collection strategy that improves contrast in astronomical imaging. An example is direct imaging of exoplanet in cameras fed by Extreme Adaptive Optics modules. 
Considering the case of Adaptive Optics (AO) observations from a ground based telescope, the ADI uses a large set of short exposures taken in pupil tracking mode,  see Figure~\ref{fig:puptrack}.
 In its original version ADI the images are median subtracted and recombined according to the actual field rotation experienced by the particular frame.
Following ADI concept, the images (star+planets) centered on the main star rotate with the sky (pupil tracking mode). The idea is subtracting the median of the images in order to wipe out the footprint of the star (the PSF), counter-rotating the residuals containing the planets.

\begin{figure}
   \begin{center}
   \begin{tabular}{c}
   \includegraphics[width=14cm]{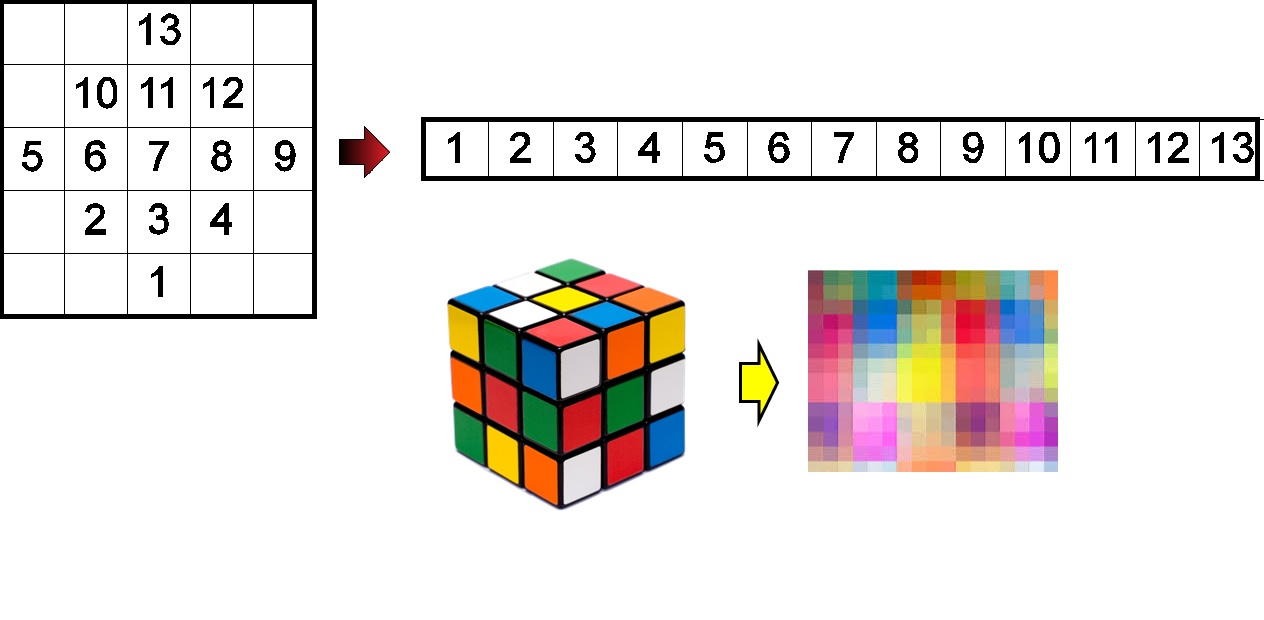}\\
   \end{tabular}
  \end{center}
   \caption[example] 
%>>>> use \label inside caption to get Fig. number with \ref{}
   { \label{fig:mat2vec} 
From each of the original frames we extract a region, and write it on a vector. The vectors are ordered on matrix.}
   \end{figure}

Using mathematical notation, we can write each image as a vector and pile up them on a matrix, see Figure~\ref{fig:mat2vec}.
The residual images ${\bf{R}}$ to be counter-rotated and summed in the ADI process can be written as:
\begin{equation}
{\bf{R}} = \sum_i {\bf{A}}_i-\hat{A} 		\\
\label{eq:ADI}
\end{equation}
Where $\hat{A}$ is the median of the available images.

\begin{figure}
   \begin{center}
   \begin{tabular}{c}
   \includegraphics[width=12cm]{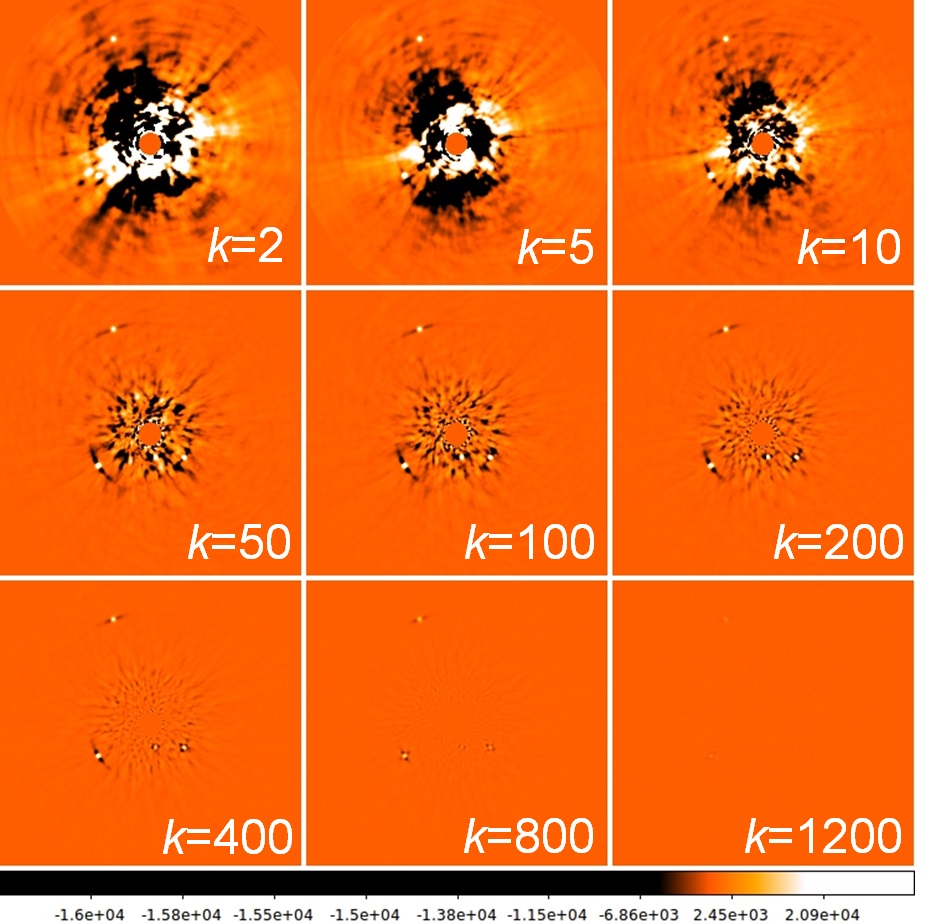}\\
   \end{tabular}
  \end{center}
   \caption[example] 
   { \label{fig:res_k} 
Increasing the rank of the matrix $\bf{A}_k$ we see that greatly improve speckles residuals. However, at the same time the flux of the companions is reduced more and more.}
   \end{figure}

The subtraction of the main dazzling object to observe the faint companion was already improved by the LOCI algorithm\cite{2007ApJ...660..770L} dividing images in subsections and obtaining,
for each subsection, a linear combination of the reference
images whose subtraction from the target image will minimize the
noise. The subtraction was further improved by the Karhunen-Lo{\`e}ve\cite{karhunen} (KL) image projection\cite{KLIP} (KLIP) algorithm. Both algorithms implement forms of Principal Component Analysis (PCA) to obtain the matrix of the residuals. We may factorize the matrix $\bf{A}\in \mathbb{R}^{m\times n}$, being $m$ the number of useful pixels on each image and $n$ the number of images, by using a general singular-value decomposition (SVD):
\begin{equation}
\bf{A} = \sum_{i=1}^n \sigma_i\bf{u_i}\bf{v_i} = \bf{U \Sigma V^T} \\
\label{eq.2}
\end{equation}
and considering just the first $k$ singular value we write the $\bf{A}$ approximation using the principal components $\bf{V}_k \in  \mathbb{R}^{m\times n}$:
\begin{equation}
\bf{A}_k  = \bf{U}_k \bf{\Sigma}_k \bf{V}_k^T \\
\label{eq.3}
\end{equation}
This factorization writes the matrix $\bf{A}$ into a matrix $\bf{A}_k$ with lower rank ($k$), projecting $\bf{A}$ on the new base $\bf{V}$. The base $\bf{V}$ is optimal, in the sense that the SVD is the factorization minimizing the residual $||\bf{R}||$: 
\begin{equation}
\bf{R}_k = \bf{A}-\bf{A}_k  \\
\label{eq.4}
\end{equation} for every possible rank $k < n$. The residual $||\cdot||$ here is the Euclidean Norm, corresponding to the {\em{rms}}. Higher the rank used larger is the effective self-subtraction of planet companion around the parent star, see Figure~\ref{fig:res_k}. 

In the following we take as data example the ADI sequences collected at LBT\cite{LBT} during the Science Demonstration Time in the October 2011 using the PISCES camera\cite{pisces} fed by the First Light Adaptive Optics (FLAO) classical AO module. The data set used is composed by $n=1396$, 5sec, H band frames of HR8799\cite{2013A&A...549A..52E} taken at LBT using PISCES+FLAO in pupil tracking mode.
   \begin{figure}
   \begin{center}
   \begin{tabular}{c}
   \includegraphics[width=8cm]{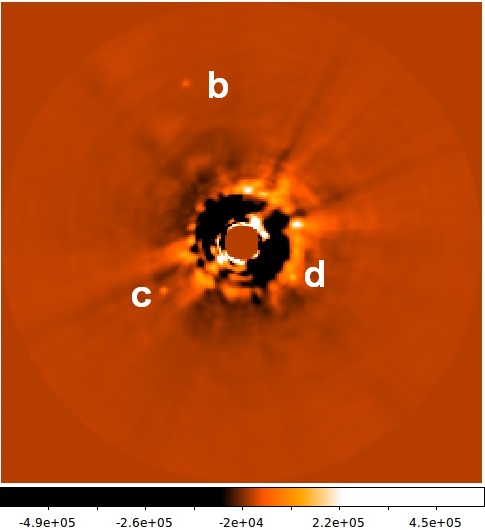}\\
   \end{tabular}
  \end{center}
   \caption[example] 
   { \label{fig:res_adi} 
The HR8799 H-band image composed by the residual of the ADI technique. Three out of four planets (b, c, d) are visible. The detection of planet "d" has poor SNR, being easily confused with speckles residuals.}
   \end{figure} 

\section{Nonnegative matrix factorization: improving Photometry and Detection}
\label{sec:nmf}
To detect a faint planet around a star we need to remove the main component image. 
A stable and known Point Spread Function (PSF) is mandatory to disentangle what is noise with respect to the true signals.
Actually ground-based adaptive optics assisted telescopes or space telescopes produce pretty stable PSF.
The Angular Differential Imaging (ADI) aims improving contrast in SCAO imaging o from Space Telescopes.
In particular it’s better achievements have been pursued on the imaging of exoplanet on NIR camera feed by an Extreme AO module.

However, in both cases long lasting- and slow evolving- speckles due to the system optical path distortion decreases the Signal to Noise Ratio (SNR) and generates false positives much more than the pure Poisson photon noise.

Take for example the case of KLIP, the methods foresees to build a KL-base for the data projection starting from a data set different than the images to be reduced: in this way aiming to reduce the effect of the self subtraction. However the faint speckles bed below the PSF is different since gravity and thermoelastic flexure are different. 
The use of the same science data set to generate the projecting space greatly improves speckles subtraction, however may fail in the detection (take the case of LOCI were the position of the companion should be known in advance) since is still dubious if low SNR objects in the reconstructed are real objects or true detections.

\begin{figure}
   \begin{center}
   \begin{tabular}{c}
   \includegraphics[width=12cm]{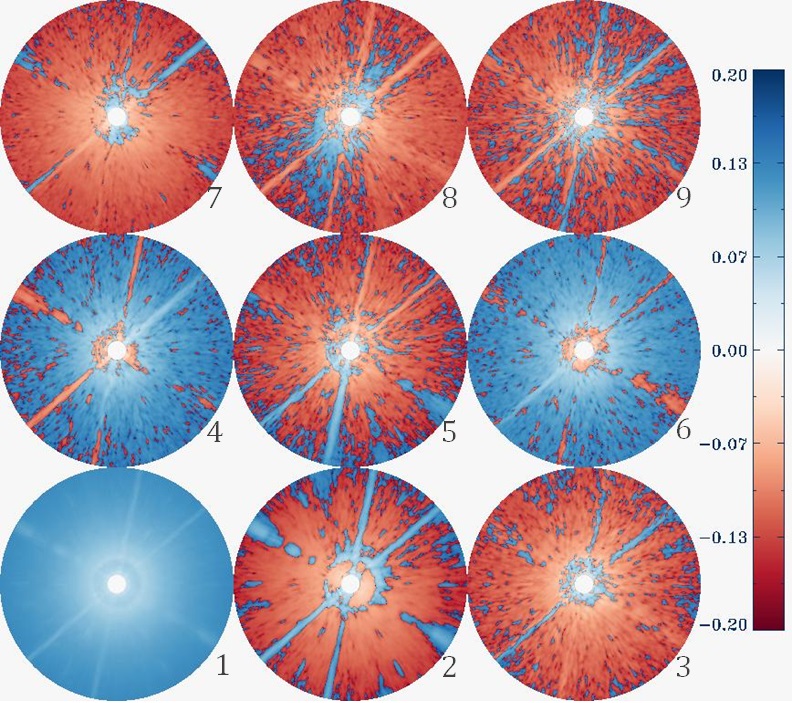}\\
   \end{tabular}
  \end{center}
   \caption[example] 
   { \label{fig:svd_fact} 
The projection space in the case of PCA is composed by positive and negative values. Only the first principal component, the "1" has physical meaning being the average of all the frames.}
   \end{figure}

Let's consider the SVD: actually the projecting base $\bf{V}$ has both positive and negative numbers and, with the exception of the first eigenmode corresponding to the average of the PSF, it has not physical meaning, see Figure~\ref{fig:svd_fact}

Our idea is to build a projecting base that is more similar to the original data set and that possibly includes a-priori information such as telescope pupil, AO telemetry, seeing value, deformable mirror (DM) characteristics etc..
Actually we foresee to write the image as it was a long exposure PSF, as the linear combination of a limited number of instantaneous PSFs, as it is in the actual image formation process.

The Nonnegative Matrix Factorization (NMF) uses only positive components, mimicking the actual composition of the long exposure images. The NMF is an approximate iterative method with less demanding computer power with respect to the PCA-like approaches, as LOCI. The PCA approach returns the best subtraction of the main star: however it is paying that in terms of self-subtraction of the faint companion that is more and more severe increasing the number of degrees of freedom used.

The NMF problem consists in the minimization of the functional
\begin{equation}
f\left(\bf{W},\bf{H}\right)=\frac{1}{2}||\bf{A}-\bf{W}\bf{H}||_F^2, {\bf{W}}\geq 0,{\bf{H}}\geq 0.
\end{equation}
Where $||\cdot||_F$ is the Froebenius norm. Actually, this problem is not convex both in {\bf{W}} and {\bf{H}}, existing multiple couples of matrices ${\bf{W}}\in \mathbb{R}^{m\times k}$  and ${\bf{H}}\in \mathbb{R}^{k\times n}$ that solve it.
Several iterative algorithms are in use to solve the problem above, such as the class of Multiplicative update algorithms\cite[Lee and Seung (2011)]{NIPS2000_1861}, that works as gradient descent algorithms, and the Alternate Least Square (ALS) that we also tried, obtaining better results.

\section{Alternate Least Square}
The Alternate Least Square algorithm already introduced in \cite[Lee and Seung (1999)]{lee99} exploits the fact that nevertheless the NMF problem is not convex in both {\bf{W}} and {\bf{H}}, it is either in {\bf{W}} or in {\bf{H}}. So, fixing one of the two, it is possible to find the unique one that solves the problem.
The idea is to use an alternate solution in {\bf{W}} and in{\bf{H}} that respects the nonnegativity condition. Starting from a random and positive {\bf{W}} we solve for {\bf{H}}:
\begin{equation}
{\bf{W}}^T{\bf{W}}{\bf{H}}={\bf{W}}^T{\bf{A}}
\end{equation}
writing {\bf{H}} as:
\begin{equation}
{\bf{H}}=\left({\bf{W}}^T{\bf{W}}\right)^{-1}{\bf{W}}^T{\bf{A}}
\end{equation}
Then we set all the negative elements of {\bf{H}} to zero and we solve for ${\bf{W}}^T$:
\begin{equation}
{\bf{H}}^T{\bf{H}}{\bf{W}}^T={\bf{H}}{\bf{A}}^T
\end{equation}
writing ${\bf{W}}^T$ as:
\begin{equation}
{\bf{W}}^T=\left({\bf{H}}^T{\bf{H}}\right)^{-1}{\bf{H}}{\bf{A}}^T
\end{equation}
and setting also in this case all negative elements of 
{\bf{W}} to zero. Repeating these operation up to convergence.

\begin{figure}
   \begin{center}
   \begin{tabular}{c}
   \includegraphics[width=12cm]{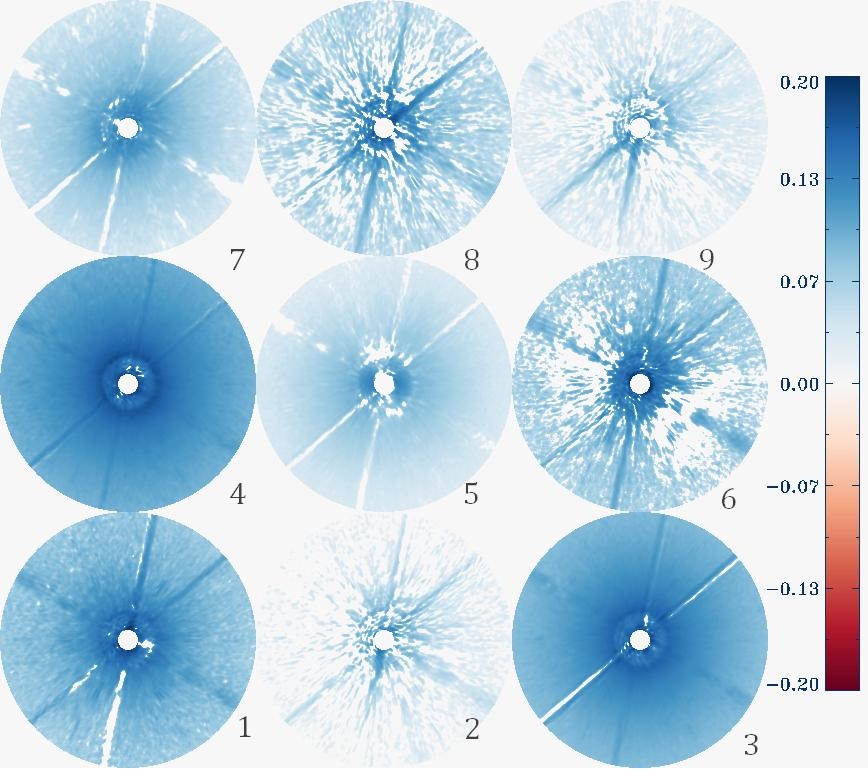}\\
   \end{tabular}
  \end{center}
   \caption[example] 
   { \label{fig:nmf_fact} 
The projection space, {\bf{H}}, in the case of NMF is composed by positive only values. All the elements have a physical interpretation, see text.}
   \end{figure}

Applying such as algorithm to our HR8799 problem, we find for the desired $k^{th}$ rank the pair of {\bf{WH}} minimizing the problem, see Figure~\ref{fig:nmf_fact}

Using this procedure we notice that a few problematic points raised: 
\begin{itemize}
\item the result depends on the initial randomization of the matrix {\bf{W}};
\item we miss a convergence criterion to avoid local minimum;
\item locking phenomenon: once one element reaches the zero it will not change anymore;
\end{itemize}
The first two points are on a strict relation: the fact that the NMF is not convex both on {\bf{W}} and {\bf{H}}, and only on {\bf{W}} or {\bf{H}} implies the convergence on local minimum, that depends on the starting point. 	

For those two problems solution exists, take for example \cite[Langville et al., 2014]{DBLP:journals/corr/LangvilleMACD14}: we tested positively a version of the SVD-centroid initialization taking as initial guess for the {\bf{H}} the positive elements of the SVD decomposition term ${\bf{V}}\in \mathbb{R}^{n\times k}$.

We applied the ALS with the SVD-centroid initialization for rank up to the 35$^{th}$, while for higher rank we noticed that the solution was not converging or converging on lower ranks with higher norm.

The projection space on which the NMF algorithm converges is a true "space of the PSF" containing the variation of the PSF on the data set: actually we found (see Figure~\ref{fig:nmf_fact}) in the  {\bf{H}} the main PSF components, such as the average PSF on frame 4, error in the centering of PSF (vibrations) on frame 3, the PSF halo and uncorrelated residuals on frame 7, spider-diffraction and residual speckles on the 2, 6 and 8, the diffraction pattern and residual within the control radius on the 5.

\begin{figure}
   \begin{center}
   \begin{tabular}{c}
   \includegraphics[width=12cm]{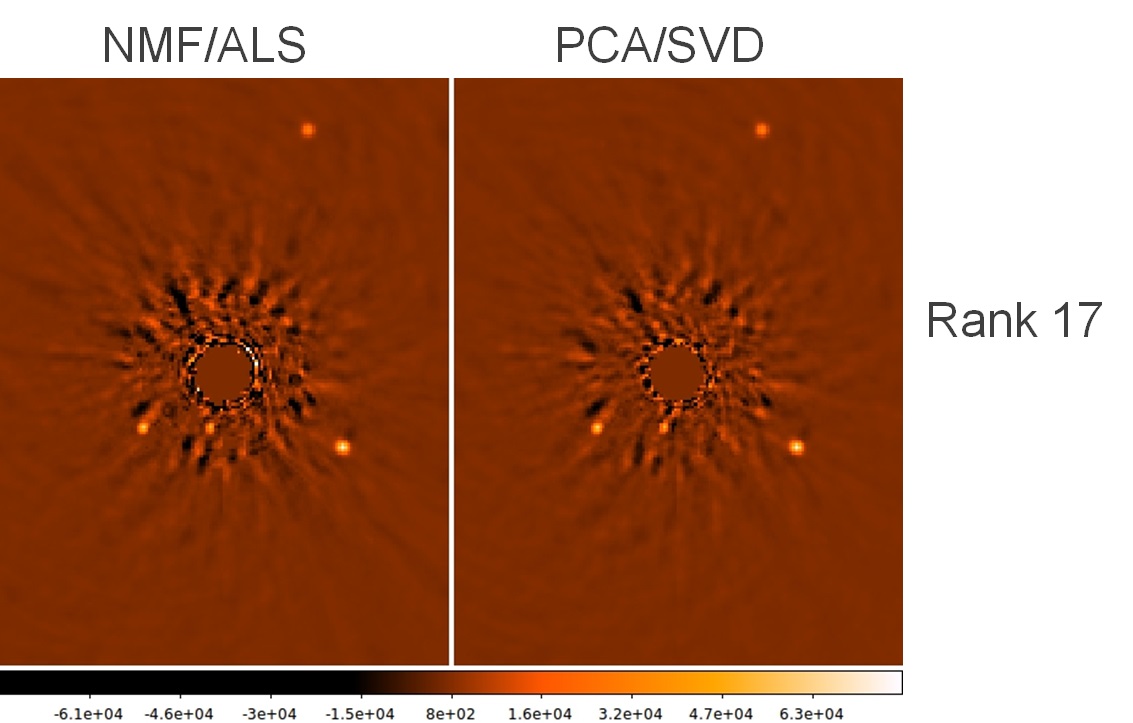}\\
   \end{tabular}
  \end{center}
   \caption[example] 
   { \label{fig:result17} 
In the frame above we show the result of NMF (right panel) and PCA-truncated SVD (left panel). The SNR of the companion are very similar, we notice a small difference in the residual speckles pattern.}
   \end{figure}

Looking into the results we found that residuals of the same rank for SVD and NMF produces very similar results, see it on Figure~\ref{fig:result17}.

\section{Further work}
Once we demonstrated the ability of the NMF to expand the images on a linear combinations of a positive space, we can include the a-priori information. A number of solutions are possible, we are converging on using auxiliary matrices orthogonal, {\bf{Q}}, or invertible {\bf{P}} in order to write the NMF problem as:
\begin{equation}
{\bf{A}}_k = {\bf{W}}{\bf{Q}}{\bf{Q}}^T{\bf{H}}   		\\
\label{eq:Q}
\end{equation}
in the case of an orthogonal matrix {\bf{Q}}, or 
\begin{equation}
{\bf{A}}_k = {\bf{W}}{\bf{P}}{\bf{P}}^{-1}{\bf{H}}   		\\
\label{eq:P}
\end{equation}
in the case of the invertible {\bf{P}}. The iterative algorithm as the ALS, in this case, would solve for {\bf{W}}{\bf{Q}} and ${\bf{Q}}^T{\bf{H}}$ instead of the simple {\bf{W}} and {\bf{H}}.
A different solution is the inclusion of penalty terms\cite{Berry2007} to enforce constraints through regularization parameters.

\section{Conclusions}
We demonstrated the ability of the NMF to solve astronomical ADI problems. The projection space on which the NMF algorithm converge is actually a "space of the PSF" containing the variation of the PSF on the data set: we are able to decompose the original data set into a linear combination of PSF elements. Moreover we explored initialization methods and propose a few ways to include a-priori constrains.

%%%%%%%%%%%%%%%%%%%%%%%%%%%%%%%%%%%%%%%%%%%%%%%%%%%%%%%%%%%%%
%%%%%%%%%%%%%%%%%%%%%%%%%%%%%%%%%%%%%%%%%%%%%%%%%%%%%%%%%%%%%
\acknowledgments     %>>>> equivalent to \section*{ACKNOWLEDGMENTS}       
The authors collaboration was established in the framework of 
"Progetto Premiale"
Adaptive Optics Made in Italy
OU 5 – Operative Unit 5 - "Analysis of adaptive Optics images".
The science data were obtained using LBT and the PISCES camera during the Science Demonstration Time in the October 2011.

%%%%%%%%%%%%%%%%%%%%%%%%\neq%%%%%%%%%%%%%%%%%%%%%%%%%%%%%%%%%%%%%
%%%%% References %%%%%

\bibliography{report}   %>>>> bibliography data in report.bib
\bibliographystyle{spiebib}   %>>>> makes bibtex use spiebib.bst

\end{document}